\documentclass[usenatbib,usegraphicx]{mn2e}
\usepackage{amssymb}

\def\Lya{Ly$\alpha\ $}

\def\HII{\hbox{H~$\scriptstyle\rm II\ $}}

\def\kms{\,{\rm {km\, s^{-1}}}}

\def\kmsmpc{\,{\rm km\,s^{-1}\,Mpc^{-1}}}

\def\ltsima{$\; \buildrel < \over \sim \;$}
\def\lsim{\lower.5ex\hbox{\ltsima}}
\def\gtsima{$\; \buildrel > \over \sim \;$}
\def\gsim{\lower.5ex\hbox{\gtsima}}

\def\spose#1{\hbox to 0pt{#1\hss}}
\def\lta{\mathrel{\spose{\lower 3pt\hbox{$\mathchar"218$}}
     \raise 2.0pt\hbox{$\mathchar"13C$}}}
\def\gta{\mathrel{\spose{\lower 3pt\hbox{$\mathchar"218$}}
     \raise 2.0pt\hbox{$\mathchar"13E$}}}

\journal{Preprint-00}

\title{The Wouthuysen-Field effect in a clumpy intergalactic medium}

\author[J. Higgins, A. Meiksin]{Jonathan Higgins, Avery Meiksin \\
SUPA\thanks{Scottish Universities Physics Alliance},
Institute for Astronomy, University of Edinburgh,
Blackford Hill, Edinburgh\ EH9\ 3HJ, UK}

\begin{document}

\maketitle

\begin{abstract}
  We show that the high optical depth of the intergalactic medium to
  \Lya photons before the Epoch of Reionization results in a
  negligible magnitude for the Wouthuysen-Field effect produced by a
  radiation source on its distant surroundings, unless (1)\ the
  scattering medium has sufficient time for the impinging resonance
  line photons to establish a steady-state frequency distribution or
  (2) the scattering gas is undergoing internal expansion or has a
  peculiar motion of tens to hundreds of $\kms$ away from the
  source. Because of intergalactic attenuation, discrete structures
  will receive only radiation from a source displaced from the
  resonance line frequency by typically hundreds to thousands of
  Doppler widths. The incident radiation must diffuse across the
  resonance line to produce a substantial scattering rate. We present
  steady-state solutions in the radiative diffusion approximation for
  the radiation field trapped in a clump of gas and show that this may
  result in an {\it enhancement} of the strength of the
  Wouthuysen-Field effect by as much as a factor of $10^6$ over the
  free-streaming (single-scattering) limit. Solutions to the
  time-dependent diffusion equation, however, show that the timescales
  required to establish a steady state will generally exceed the
  lifetime of the sources, resulting in a substantially reduced
  scattering rate. In the presence of internal expansion, a steady
  state may be established as photons are redshifted across the
  resonance line and into the red wing, and significant enhancement in
  the scattering rate over the free-streaming limit may again be
  produced. Alternatively, a substantial scattering rate may arise in
  systems with a peculiar motion away from the source that redshifts
  the received radiation into the resonance line centre. As a
  consequence, at epochs $z\lta30$, when collisional decoupling of the
  hyperfine structure of hydrogen from the Cosmic Microwave Background
  is small except in dense regions, and prior to the establishment of
  any large-scale diffuse radiation field of resonance line photons,
  the 21cm signature from the Intergalactic Medium produced by the
  Wouthuysen-Field effect will in general trace the peculiar velocity
  field of the gas in addition to its density structure.
\end{abstract}


\section{Introduction}

Following the Recombination Era at $z\simeq1100$, the baryons produced
in the Big Bang were largely neutral. By $z\gta6$, the spectra of high
redshift Quasi-Stellar Objects (QSOs) show that the hydrogen had
become ionized \citep{Becker01}. The history of structure formation in
the wide redshift expanse between these epochs, when there were few if
any sources of radiation, is largely unknown. In principle, 21cm
emission from intergalactic hydrogen could reveal the evolution of
structure in the baryons during these cosmic ``Dark Ages.''  Because
the baryons are cold, they would closely trace the evolution of the
dark matter, so that 21cm imaging could trace the growth of structure
following the Recombination Era \citep{1979MNRAS.188..791H,
1990MNRAS.247..510S}. Because of strong coupling between the hyperfine
structure of the hydrogen and the Cosmic Microwave Background (CMB),
however, a mechanism that decouples the spin temperature of the
hydrogen hyperfine structure from the CMB temperature must be active,
otherwise the hydrogen is rendered invisible:\ it absorbs and re-emits
the CMB radiation at the same rate, leaving it indistinguishable from
the CMB. At redshifts $z>30$, collisions between hydrogen and
electrons and other hydrogen atoms are adequate to begin decoupling
the spin temperature from the CMB \citep{1990MNRAS.247..510S,
MMR97}. Except in dense regions, however, at later times collisional
decoupling is inadequate.

Once the first sources of radiation begin to turn on, the scattering
of \Lya photons by the neutral hydrogen offers an alternative means of
decoupling, through the Wouthuysen-Field effect (WFE)
\citep{1952AJ.....57R..31W, 1958PROCIRE.46..240F}. As sources turn on
in sufficient number to begin reionizing the Intergalactic Medium
(IGM), they will produce a combined intensity of \Lya photons
sufficient for coupling the spin temperature to the light temperature
of the \Lya photons rather than the CMB, rendering the intergalactic
hydrogen visible against the CMB in either emission or absorption
\citep{MMR97}.

The discovery of the End of the Dark Ages (EDA) and the onset of the
Epoch of Reionization (EoR) has become one of the paramount goals of a
new generation of radio telescopes, such as the LOw Frequency Array
(LOFAR)\footnote{www.lofar.org}, the Murchison Widefield Array (MWA)
\footnote{www.haystack.mit.edu/ast/arrays/mwa}, the Primeval Structure
Telescope/21 Centimeter Array (PaST/21CMA)
\footnote{web.phys.cmu.edu/$\sim$past}, the Precision Array to Probe
EoR (PAPER)\footnote{astro.berkeley.edu/$\sim$dbacker/eor}, and a
possible Square Kilometre Array (SKA)\footnote{www.skatelescope.org}.
Reviews of this rapidly growing area are provided by
\citet{2001ARA&A..39...19L}, \citet{2006ARA&A..44..415F} and
\citet{2006PhR...433..181F}.

\citet{MMR97} estimated the \Lya collision rate $P_\alpha$ that drives
the WFE as the integrated intensity from cosmologically distributed
sources, assuming that photons blueward of \Lya emitted from a source
will contribute their full amount to $P_\alpha$ once they redshift
into the local \Lya resonance. In fact, this provides a lower limit
(assuming photons are not destroyed by, eg, dust absorption). The
multiple scattering of resonance line photons will in general enhance
the radiation field. For pure Doppler scattering,
\citet{1959ApJ...129..536F} argued the rate is enhanced by the number
of scatters a photon undergoes before being randomly scattered
sufficiently redward of line centre to escape. The estimate neglected
scattering in the Lorentz wings, however. It also assumed that the IGM
takes part in the homogeneous and isotropic expansion of the
Universe. In fact the IGM is clumpy, with structures breaking away
from the cosmological expansion. Efforts are underway to estimate the
radiation field and scattering rate in a cosmological context through
Monte Carlo simulations of the scattering of resonance line photons in
an inhomogeneous medium \citep{2007ApJ...670..912C,
2007arXiv0712.1159P, 2007A&A...474..365S}.  In this paper, we use
analytical means to explore some of the consequences of intergalactic
structure formation for the WFE as a means of generating an
intergalactic 21cm signature.

In the next section, we discuss the optical depth of the IGM to
resonance line photons and the implications for the scattering rate.
Approximate steady-state solutions to the radiative transfer equation
are derived in \S~\ref{sec:discrete}, and time-dependent solutions in
\S~\ref{sec:evol}. A discussion of the solutions and applications are
provided in \S~\ref{sec:disc}, along with a summary of our
conclusions.

\section{The WFE in the intergalactic medium}
\label{sec:IGM}

The optical depth through a homogeneous and isotropic expanding IGM of
a photon emitted by a source at redshift $z_S$ and received at
redshift $z$ at frequency $\nu>\nu_0$, where $\nu_0$ is the resonance
line frequency, is \citep{1959ApJ...129..536F, 1965ApJ...142.1633G}
\begin{equation}
\tau_{\nu} = \sigma\int_z^{z_S}\,dz^\prime \frac{dl_p}{dz^\prime}
n_l(z^\prime)\varphi_V\left(a,\nu\frac{1+z^\prime}{1+z}\right),
\label{eq:taunu}
\end{equation}
where $n_l(z^\prime)$ is the number density of scattering atoms in the
lower level at epoch $z^\prime$, $\sigma=\pi e^2 f_{lu}/(m_e
c)\simeq0.0110\,{\rm cm^2\,Hz}$ is the total resonance line cross
section, where $f_{lu}\simeq0.4162$ is the upward oscillator strength
for hydrogen Ly$\alpha$, $\varphi_V(a,\nu)$ is the Voigt line profile
normalized to $\int\, d\nu\varphi_V(a,\nu)=1$, $a\simeq0.0472T^{-1/2}$
is the ratio of the decay rate to the Doppler width $\Delta\nu_D=\nu_0
b/c$, where $b=(2k_{\rm B}T/m_{\rm H})^{1/2}$ is the Doppler parameter
for hydrogen gas at temperature $T$ and $c$ is the speed of light, and
$l_p$ is the proper path length. In the Lorentz wing, expressed as a
function of $x=(\nu-\nu_0)/\Delta\nu_D$, the dimensionless Voigt
profile $\phi_V(a,x)=(\Delta\nu_D)\varphi_V(a,\nu)$ is
well-approximated as $\phi_V(a,x)\simeq a/(\pi x^2)$. The differential
proper line element evolves according to
$dl_p/dz\simeq(c/H_0)\Omega_m^{-1/2} (1+z)^{-5/2}$ in a flat universe
at redshifts for which $\Omega_m(1+z)^3$ dominates the contribution
from the vacuum energy, where $\Omega_m$ is the ratio of the total
mass density to the critical Einstein-deSitter density, and where
$H_0=100h\kmsmpc$ is the Hubble constant today.

\begin{figure}
\includegraphics[width=3.3in]{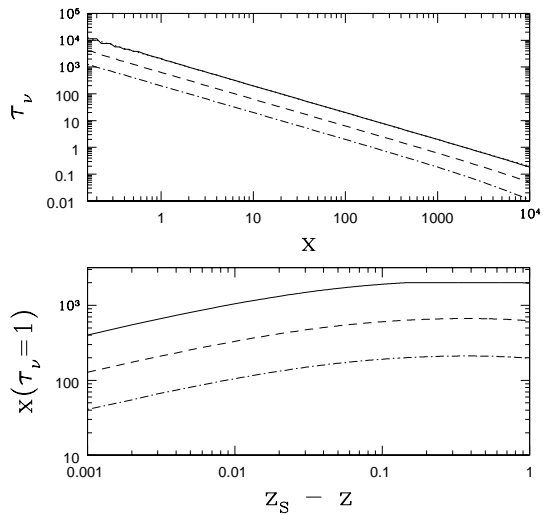}
\caption{(Upper panel)\ The optical depth of photons emitted by a
source at redshift $z_S$ and received at the frequency $\nu$ near the
resonance line frequency $\nu_0$ of \Lya at redshift $z$. The
frequency is expressed as $x=(\nu-\nu_0)/\Delta\nu_D$. The curves are
for $T=10$~K (solid), 100~K (dashed) and 1000~K (dot-dashed). Also
shown is the approximation $\tau_\nu\simeq x_1/x$ for the $T=10$~K
model (dotted line, nearly coinciding with the solid line). (Lower
panel)\ The value of $x$ at which $\tau_\nu=1$ for $z_S - 1 < z <
z_S$. The curves are labelled as in the upper panel.
}
\label{fig:taunu}
\end{figure}

Photons emitted by the source at a frequency
$\nu_e>\nu_0(1+z_S)/(1+z)$ will scatter in the blue Lorentz wing
(except for those for which the inequality is nearly an equality, in
which case they will scatter through the Doppler core). In the blue
wing the optical depth due to \Lya\ scattering by neutral hydrogen is
\begin{eqnarray}
\tau_{\nu}&\simeq&1.75\times10^5\frac{cn_l(0)}{H_0\Omega_m^{1/2}}
\frac{(1+z)^{3/2}}{\nu_0^2{y^{3/2}}}
\Biggl\{\left[\frac{y^{1/2}}{y-1}-\frac{u^{1/2}}{u-1}\right]\nonumber\\
&&\phantom{\Biggl\{}+\frac{1}{2}\log\left[\frac{(1-u^{1/2})(1+y^{1/2})}
{(1-y^{1/2})(1+u^{1/2})}\right]\Biggr\},
\label{eq:taununo}
\end{eqnarray}
where $y=\nu/\nu_0$ and $u=y(1+z_S)/(1+z)$, and all quantities are
assumed to be in cgs units. For $0<xb/c<<(z_S-z)/(1+z)<<1$, the
expression simplifies to
\begin{equation}
\tau_{\nu}\approx \frac{x_1}{x},
\label{eq:taunu_ap}
\end{equation}
where
\begin{equation}
x_1=\frac{a}{\pi}\frac{\sigma
c}{\nu_0}\frac{n_l(0)}{H_0\Omega_m^{1/2}}(1+z)^{3/2}
\approx214h^{-1}(1+z)^{3/2}T_{\rm IGM}^{-1/2}
\label{eq:x1}
\end{equation}
for a mean line-of-sight IGM temperature $T_{\rm IGM}$ for a universe
with $\Omega_m=0.3$ and baryon density $\Omega_bh^2=0.022$
\citep{2006ApJ...649L..61O, Spergel07}. As an illustration,
$\tau_\nu(x)$ for a source at $z_S=8$ is shown at $z=7$ in
Fig.~\ref{fig:taunu} (upper panel). The optical depth is extremely
high until $\nu$ is well displaced from the line centre. The values of
$x_1$, at which $\tau_\nu=1$, are shown in the lower panel of
Fig.~\ref{fig:taunu} for a range of redshifts.

Photons emitted by the source with frequencies between
$\nu_0<\nu_e<\nu_0(1+z_S)/(1+z)$ will pass through the resonance line
frequency en route to the gas at $z$. Upon passing through the
resonance line, the photons will essentially be completely scattered
out of the line of sight. As a consequence, essentially no photons
will be received in the frequency range $\nu_0(1+z)/(1+z_S)<\nu<\nu_0$
(the Gunn-Peterson effect). A general expression for the optical depth
through the Doppler core is provided in Appendix~\ref{ap:taucore}.

Photons emitted sufficiently redward of $\nu_0$ to avoid the Doppler
core may still be scattered in the Lorentz red wing of the surrounding
hydrogen. This will produce a halo of scattered \Lya photons around a
central source like a QSO \citep{1999ApJ...524..527L}. For frequencies
$x<0$ near line centre, \citet{1999ApJ...524..527L} show that the
photons will build up an energy density as they diffuse through the
surrounding IGM of the form $u_\nu\sim (-x)^{-9/2}e^{-[r/r_c(x)]^2}$,
where $r_c(x)=(-x)^{1/2}(2/3)[(b/H)\lambda_{\rm mfp}(x)]^{1/2}$ scales
like the harmonic mean between the distance over which Hubble
expansion produces a velocity difference matching the Doppler
parameter $b$ and the scattering mean free path of a resonance line
photon in the Lorentz wing
\begin{equation}
\lambda_{\rm mfp}(x)\simeq\frac{\pi\Delta\nu_D}{n_l\sigma a}x^2,
\label{eq:mfp}
\end{equation}
where $n_l$ is the number density of atoms in the lower level. The
energy density is exponentially suppressed for $x\rightarrow0^{-}$
at a fixed position, resulting in a negligible \Lya scattering rate
except very near the QSO. Numerically,
\begin{equation}
r_c(x)\simeq0.6\,{\rm pc}(-x)^{3/2}h^{-1/2}T_{\rm
IGM}^{3/4}\Omega_m^{-1/4} \left(\frac{1+z}{9}\right)^{-9/4},
\label{eq:rc}
\end{equation}
so that a substantial \Lya scattering rate is confined to a small
region around the source. The region is so small in fact, that the
assumption that the surrounding IGM takes part in the Hubble flow will
not be valid. More likely the gas will have been ionized and violently
disturbed by the QSO itself.

Finally, photons emitted sufficiently redward of $\nu_0$ at the source
to escape scattering in the Lorentz wing will pass freely through the
IGM, where they will be received at redshift $z$ at frequencies
$x<-x_0$, where
\begin{equation}
x_0=\frac{c}{b_{\rm IGM}}\frac{z_S-z}{1+z_S}\simeq
2.3\times10^6T_{\rm IGM}^{-1/2}\frac{z_S-z}{1+z_S}.
\label{eq:x0}
\end{equation}

{\it The vast majority of photons from the source emitted blueward of,
but near, the resonance line frequency never arrive near the line
centre, having been scattered out of the path along the way.} The
resonance line radiation impinging on the absorption
site\footnote{Strictly speaking, the photons are scattered. The region
at which the scattering rate is computed is referred to for simplicity
as the absorption site or the absorber.}  from the source scatters in
the \Lya resonance line at the rate per neutral atom in the lower
state
\begin{equation}
P_l=\sigma c\int_0^\infty\,d\nu \varphi_V(a,\nu)
\frac{u_\nu}{h_{\rm P}\nu},
\label{eq:Pl}
\end{equation}
where $h_{\rm P}$ is Planck's constant and $u_\nu$ is the local
specific energy density of photons received at $z$ from the source at
$z_S$ with specific luminosity $L_\nu$
\begin{equation}
u_\nu=\frac{1}{c}\frac{L_\nu}{4\pi r_L^2}\exp(-\tau_\nu)
\label{eq:unu}
\end{equation}
where $r_L$ is the luminosity distance between the source and the gas
at redshift $z$. The scattering rate assuming none of the photons have
been scattered out of the line of sight en route to the absorption
site (the free-streaming limit) is
\begin{equation}
P_l(\tau_\nu=0)\simeq\sigma\frac{L_\nu/h_{\rm P}\nu}{4\pi r_L^2}.
\label{eq:Pl0}
\end{equation}
The ratio of the scattering rate for $\tau_\nu\simeq x_1/x$ to the
free-streaming value is then given by
\begin{eqnarray}
\frac{P_l(\tau_\nu)}{P_l(\tau_\nu=0)}&\simeq&
\frac{\int_0^{\infty}dx a \pi^{-1}x^{-2}\exp(-x_1/x)}
{\int_{-\infty}^{\infty}dx\phi_V(a,x)}=\frac{a}{\pi x_1}\nonumber\\
&\simeq&
7.0\times10^{-5}h(1+z)^{-3/2}\left(\frac{T_{\rm IGM}}{T_a}\right)^{1/2},
\label{eq:Plratio}
\end{eqnarray}
where $T_a$ is the temperature of the absorbing gas. The low value
poses a fundamental problem to the effectiveness of the WFE as a means
of decoupling the spin state of the gas from statistical equilibrium
with the CMB. {\it Only if photons have sufficient time to diffuse
across the line centre will the scattering of \Lya photons be an
effective means of decoupling the spin temperature from the CMB in
comoving objects.}

\section{The WFE in discrete objects}
\label{sec:discrete}

\subsection{Steady-state solutions without atomic recoil}
\label{subsec:ss}

The assumption of a homogeneous and isotropic universe, while a fair
approximation over large scales, breaks down on small scales:\ the IGM
is clumpy at the redshifts from which the 21cm signal is expected
\citep{2000ApJ...528..597T, 2004ApJ...608..611G}. For a sufficiently
short mean free path, photons will be trapped within a discrete system
and diffuse in frequency and space as they scatter within the
system. In this section, we consider the energy distribution the
radiation will reach given adequate time to relax to a steady state.

For a \Lya\ photon to become trapped within an absorber of lengthscale
$L_a$, the mean free path must minimally satisfy $\lambda_{\rm
mfp}<L_a$, for which $\tau_\nu=L_a/\lambda_{\rm mfp}>1$. As the
optical depth in the wing may be expressed as
$\tau_\nu=\pi^{-1/2}(a\tau_0)x^{-2}$, where $\tau_0=n_lL_a\pi^{-1/2}
\sigma (\Delta\nu_D)^{-1}$ is the optical depth at the line centre,
this corresponds to photons escaping with
$x>\pi^{-1/4}(a\tau_0)^{1/2}$ \citep{1962ApJ...135..195O}. Radiative
transfer solutions for sources embedded within a slab suggest this may
be somewhat too restrictive for systems with very high line centre
optical depths, in which case the photons escape through spatial
diffusion rather than frequency diffusion. Steady-state solutions in
the diffusion approximation for scattering in the Lorentz wings in
very high optical depth systems show that photons escape the slabs
with frequencies typically on the order of $(a\tau_0)^{1/3}$
\citep{1972ApJ...174..439A,1973MNRAS.162...43H,1990ApJ...350..216N}. An
escape frequency of $x_{\rm esc}=x_*(a\tau_0)^{1/3}$ corresponds to a
mean free path smaller than the thickness of the slab by the factor
$f=x_*^2\pi^{1/2}(a\tau_0)^{-1/3}
\simeq0.013x_*^2T_a^{1/3}N_{19}^{-1/3}$, where $T_a$ is the
temperature of the absorber and $N_{19}$ is the column density in
units of $10^{19}\,{\rm cm^{-2}}$. (This corresponds to a region
24~kpc across for the mean intergalactic hydrogen density at $z=8$,
and smaller for overdense systems.) As $f$ depends only weakly on
$T_a$ and $N_{\rm HI}$, while $x_*$ depends on $a\tau_0$ and the
radiation source parameters of any particular problem, we express the
escape frequency simply as
\begin{equation}
x_{\rm esc}\simeq395(10f)^{1/2}N^{1/2}_{19}T_a^{-1/2},
\label{eq:xesc}
\end{equation}
noting that $f$ will in general vary with the depth within the
absorber, and must be solved for according to each particular
configuration, but will be of order $0.01-1$ for applications we
consider. Comparison of Eq.~(\ref{eq:xesc}) with Eq.~(\ref{eq:x1}) for
$x_1$ shows that photons with $\tau_\nu<1$ in the IGM will generally
not be trapped within a structure comoving with the expansion of the
Universe unless the structure has a neutral hydrogen column density
$N_{19}>N_{\rm 19, crit}$, where
\begin{equation}
N_{\rm 19, crit}\simeq0.29h^{-2}(10f)^{-1}(1+z)^3
\left(\frac{T_a}{T_{\rm IGM}}\right).
\label{eq:N19crit}
\end{equation}

If $x_1<x_{\rm esc}$, the absorber will behave as a photon bucket,
allowing the energy density of the radiation field at frequencies
$x<x_{\rm esc}$ to build up with time as radiation from the source
becomes trapped. As the photons scatter within the absorber, they will
diffuse in frequency and space. Because photons with $\vert
x\vert>x_{\rm esc}$ will escape, the energy density of the radiation
trapped within the absorber will reach a steady state when the rate of
incoming photons balances the rate of escape. In principle, the energy
density that develops could produce a substantial scattering rate,
even exceeding the rate estimated assuming no photon losses from the
source due to scattering by the intervening IGM.

The depth to which photons at frequency $x_1$ penetrate an absorber
depends on the optical depth of the absorber. If it is optically
thin at $x_1$, the photons will stream through. If optically thick,
the photons will scatter near the surface of the absorber. Photons
blueward of $x_1$, however, will penetrate more deeply, as the cross
section diminishes like $1/x^2$. In general, photons will be injected
at varying layers within the absorber, with bluer photons injected at
increasing depths.

We would like to estimate the evolution of the radiation field within
an absorber. No solution to this problem exists in the literature. The
full radiative transfer problem through the slab is an involved one,
as the radiation field will vary with both the optical depth through
the slab and with frequency. It is particularly important in our
application to obtain the solution across the Doppler core to estimate
the \Lya\ scattering rate. Instead of solving the full problem, we
seek approximate solutions in which we treat the scattered radiation
field as locally isotropic, neglecting the spatial diffusion of the
radiation through the absorber. We show in Appendix~\ref{ap:slab} that
the resulting radiative transfer equation in the diffusion
approximation is formally identical to the equation of
\citet{1973MNRAS.162...43H} for the problem of a slab with a uniform
distribution of spectrally flat sources. The solution we obtain
neglecting the spatial diffusion of the radiation agrees well with
Harrington's solution which includes both spatial and frequency
diffusion. A detailed quantitative description will require a more
exact solution to the problem, but our approximate approach allows us
to explore qualitatively several effects of interest. We would like to
estimate the energy density that may be achieved as a function of
$x_{\rm esc}$. We also wish to examine the effect of internal motions
on the radiation energy density across the line centre and the
possible role of atomic recoil. Lastly, we would like to estimate the
time it takes the radiation field to establish a steady state. All of
these effects extend well beyond those explored by existing slab
solutions.

We estimate the radiation density that builds up within an absorber by
assuming a steady state between the rate of photons injected into the
absorber and the rate of diffusion of the photons across the escape
frequency, imposing the boundary condition $u(x)=0$ for $\vert x\vert
> x_{\rm esc}$, where $u(x)=\Delta\nu_D u_\nu$. We assume isotropic
scattering within a homogeneous absorber, and use the diffusion
approximation to describe the resonance line scattering of the
photons. In equilibrium,
\begin{equation}
\frac{1}{2}\frac{d}{dx}\left[D(a,x)\frac{dn(x)}{dx}\right]=-\tilde S(x),
\label{eq:FPDA}
\end{equation}
where $n(x)=\Delta\nu_D n_\nu$, $n_\nu=u_\nu/h_{\rm P}\nu$, $\tilde
S(x)=(\Delta\nu_D)^2S(\nu)/(n_l c\sigma)$, where $S(\nu)$ is a source
function, and $D(a,x)$ is the diffusion coefficient. The second order
moment of the frequency redistribution function gives for the
diffusion coefficient $D(a,x)=\phi_V(a,x)+(1/3)d^2\phi_V(a,x)/dx^2$
\citep{1994ApJ...427..603R}. The derivation of the diffusion equation,
however, does not conserve photon number to the order of the
approximation. As a consequence, the diffusion coefficient is left
ambiguous. \citet{1994ApJ...427..603R} advocate adopting
$D(a,x)=\phi_V(a,x)$ for its simplicity, and we shall do so here.

\begin{figure}
\includegraphics[width=3.3in]{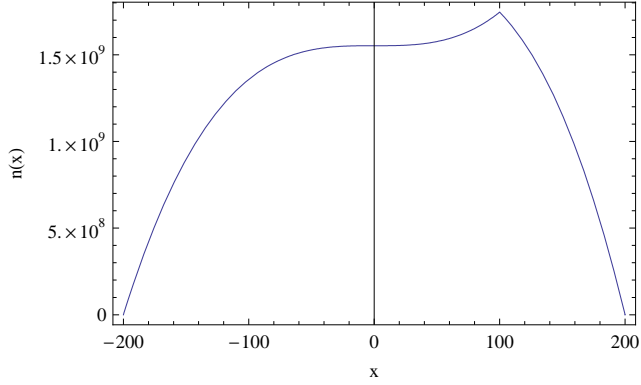}
\caption{The steady-state distribution of \Lya\ resonance line photons
for a source function $\tilde S(x)=\delta_D(x-x_{\rm inj})$ with
$x_{\rm inj}=100$, allowing photons with $\vert x\vert>x_{\rm
esc}=200$ to escape, as given by the diffusion
approximation, Eq.~(\ref{eq:FPDA}), with $D(a,x)=\phi_V(a,x)$.
An absorber temperature $T_a=100$~K is assumed.
}
\label{fig:nxFPDA-FPA}
\end{figure}

The source $S(\nu)$ represents the rate at which photons received from
the source are incident on the absorber at $x<x_{\rm esc}$, where they
are trapped and diffuse in frequency. It is given by the scattering
emissivity
\begin{equation}
S(\nu) = n_lc\sigma\int_0^\infty\, d\nu^\prime
R(\nu^\prime,\nu)n_{\nu^\prime},
\label{eq:Snu}
\end{equation}
where $R(\nu^\prime,\nu)$ is the photon redistribution function for
resonance line scattering \citep{1978stat.book.....M}. Since
scattering in the wings is coherent,
$R(\nu^\prime,\nu)\simeq\varphi_V(a,\nu^\prime)\delta_D(\nu^\prime-\nu)$,
where $\delta_D$ is the Dirac $\delta$-function. Then
\begin{eqnarray}
S(\nu)&\simeq& n_lc\sigma\varphi_V(a,\nu)n_\nu\nonumber\\
&\simeq& n_l\sigma a\pi^{-1}(\Delta\nu_D)^{-1}\frac{L_\nu/h_{\rm
P}\nu}{4\pi r_L^2} x^{-2}\exp(-x_1/x).
\label{eq:SnuLnu}
\end{eqnarray}
The dependence on $x$ results in a sharp peak at $x=x_1/2$, so that
the source $\tilde S(x)$ is well approximated by
\begin{eqnarray}
\tilde S(x)&\simeq& \frac{a}{\pi c}\frac{\Delta\nu_D}{x_1}
\frac{L_\nu/h_{\rm P}\nu}{4\pi r_L^2}\delta_D(x-x_{\rm inj})\nonumber\\
&=&A\delta_D(x-x_{\rm inj})
\label{eq:Sx}
\end{eqnarray}
where the normalization $A=2.5\times10^{-6}h(1+z)^{-3/2}T_{\rm
IGM}^{1/2}[(L_\nu/h_{\rm P}\nu)/4\pi r_L^2]$ (with all quantities
expressed in cgs units), matches the source rate of
Eq.~(\ref{eq:SnuLnu}) integrated over all frequencies. Note that the
dependence on the absorber temperature cancels. Here, the frequency at
which photons are injected into the absorber is represented generally
as $x_{\rm inj}$ to allow for any peculiar motion of the absorber,
which could shift the peak of the received radiation either redward or
blueward, and to allow for scattering within the absorber which will
shift the peak of the radiation field penetrating to deeper layers
further toward the blue. For a comoving absorber, $x_{\rm inj}=x_1/2$
near the surface. The solution to Eq.~(\ref{eq:FPDA}) using the wing
approximation for $\phi_V(a,x)$ is
\begin{equation}
n(x) = \cases{\frac{\pi}{3a}A\left(1-
\frac{x_{\rm inj}^3}{x_{\rm esc}^3}\right)
(x_{\rm esc}^3+x^3) & ; $-x_{\rm esc}<x<x_{\rm inj}$\cr
\frac{\pi}{3a}A\left(1+\frac{x_{\rm inj}^3}{x_{\rm esc}^3}\right)
(x_{\rm esc}^3-x^3) &; $x_{\rm inj}<x<x_{\rm esc}.$\cr}
\label{eq:nxFPDAss}
\end{equation}
The solution is shown in Fig.~\ref{fig:nxFPDA-FPA}. We find that
adopting $D(a,x)=\phi_V(a,x)+(1/3)d^2\phi_V(a,x)/dx^2$ produces a
nearly identical solution, agreeing to within a fraction of a percent
for $x_{\rm inj}$ well out of the core. Numerically integrating the
diffusion equation using the full Voigt profile produces a solution
that also differs negligibly from Eq.~(\ref{eq:nxFPDAss}).

The ratio of the collision rate $P_l$, given by using
Eq.~(\ref{eq:nxFPDAss}) in Eq.~(\ref{eq:Pl}), to the free-streaming
collision rate $P_l(0)$ of Eq.~(\ref{eq:Pl0}) is then
\begin{eqnarray}
\frac{P_{l, {\rm ss}}}{P_l(0)}&\simeq&\frac{1}{x_1}
\frac{a}{\pi A}\int_{-\infty}^{\infty}dx\,\phi_V(a,x)n(x)\nonumber\\
&\simeq&\frac{1}{3x_1}
\left[\left(1-\frac{x_{\rm inj}^3}{x_{\rm esc}^3}\right)x_{\rm esc}^3+
\frac{a}{\pi}\left(3x_{\rm inj}^2-x_{\rm esc}^2-2\frac{x_{\rm inj}^3}
{x_{\rm esc}}\right)\right]\nonumber\\
&\simeq&\frac{1}{3x_1}\left(1-\frac{x_{\rm inj}^3}{x_{\rm esc}^3}\right)
x_{\rm esc}^3\nonumber\\
&\simeq&9.6\times10^4h\left[\frac{(10f)N_{19}}{(1+z)T_a}\right]^{3/2}
T_{\rm IGM}^{1/2}\left(1-\frac{x_{\rm inj}^3}{x_{\rm esc}^3}\right),
\label{eq:Plratioss}
\end{eqnarray}
where we have used the approximation Eq.~(\ref{eq:phiVap}) below for
the Voigt profile with $x_{\rm esc}>x_{\rm inj}\gg x_m $ assumed, and
neglected the terms following the leading. The enhancement may be
substantial. For $N_{19}=N_{19, \rm crit}$, $P_{l, {\rm
ss}}/P_l(0)\simeq 1.5\times10^4h^{-2}(1+z)^3T_{\rm IGM}^{-1} (1-x_{\rm
inj}^3/x_{\rm esc}^3)$, so that even for a warm IGM temperature of
$T_{\rm IGM}=1000$~K a substantial boost in the scattering rate will
result.

\begin{figure}
\includegraphics[width=3.3in]{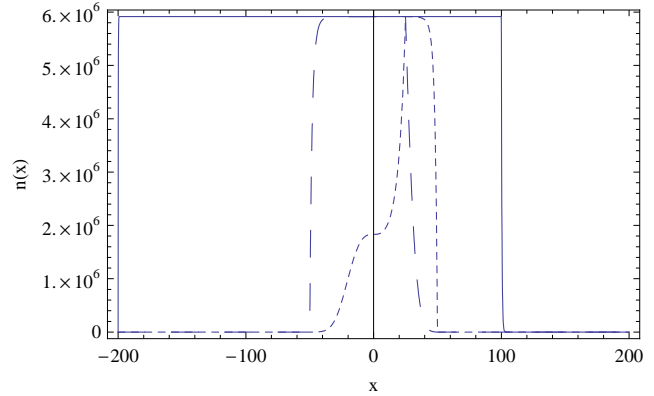}
\caption{The steady-state distribution of \Lya\ resonance line photons
for a source function $\tilde S(x)=\delta_D(x-x_{\rm inj})$ with
$x_{\rm inj}=100$ in an absorber with temperature $T_a=100$~K,
allowing photons with $\vert x\vert>x_{\rm esc}=200$ to escape and
including the effects of internal motions with a dimensionless
velocity gradient $\gamma=1.7\times10^{-7}$ (solid line), representing
expansion. Virtually all the injected photons are redshifted, escaping
for $x<-x_{\rm esc}$. The solution for $x_{\rm inj}=25$ and $x_{\rm
esc}=50$ for $\gamma=1.7\times10^{-7}$ is also shown (long-dashed
line), showing a weak tail at $x>x_{\rm inj}$. The corresponding case
with contraction, $\gamma=-1.7\times10^{-7}$ (short-dashed line) shows
a more complex behaviour as contraction curtails the redward
diffusion.
}
\label{fig:nxFPDAg}
\end{figure}

The solution Eq.~(\ref{eq:nxFPDAss}) assumes velocities internal to
the cloud are negligible. To estimate the effect internal motion may
have on the frequency distribution of the photons, we approximate the
motion as an isotropic expansion (or contraction) within the cloud
characterised by a uniform velocity gradient $H_a$ (which in general
will not be the same as the Hubble parameter). In the diffusion
approximation, Eq.~(\ref{eq:FPDA}) is modified to
\citep{1994ApJ...427..603R}
\begin{equation}
\frac{1}{2}\frac{d}{dx}\left[2\gamma n(x)+D(a,x)\frac{dn(x)}{dx}\right]=
-\tilde S(x),
\label{eq:FPDAg}
\end{equation}
where $\gamma=\nu_0 H_a/(\sigma n_l c)$ is $(c/b)$ times the ratio of
the scattering time of the photons at line centre to the expansion
time $H_a^{-1}$ of the gas. For this solution, the steady state
applies to the comoving number density of photons per frequency
\citep{1994ApJ...427..603R}. Overdense structures are typically
contracting, but the internal motions along a filament may expand
along the filament in regions \citep{ZMAN98}. For an overdensity ten
times the cosmic mean at $z=8$, an internal velocity gradient of
$100\kms$ over 100~kpc corresponds to
$\gamma\simeq1.7\times10^{-7}$. Contraction by the same magnitude
corresponds to $\gamma\simeq-1.7\times10^{-7}$.

The solution for the source $\tilde S(x)=A\delta_D(x-x_{\rm inj})$ is
\begin{equation}
n(x) = \cases{g_\gamma(x,-x_{\rm esc}) & ; $-x_{\rm esc}<x<x_{\rm inj}$\cr
g_\gamma(x,-x_{\rm esc})-g_\gamma(x,x_{\rm inj})
&; $x_{\rm inj}<x<x_{\rm esc},$\cr},
\label{eq:nxFPDAssg}
\end{equation}
where $g_\gamma(x,y)=(A/\gamma)f_\gamma(x_{\rm esc},x_{\rm inj})
f_\gamma(x,y)/f_\gamma(x_{\rm esc},y)$ and
$f_\gamma(x,y)=1-\exp[-\frac{2\pi\gamma}{3a}(x^3-y^3)]$. The solution
is extremely sensitive to $\gamma x_{\rm esc}^3/a$. For
$\gamma\gg\gamma_{\rm crit}\equiv3a/(2\pi x_{\rm esc}^3)$, a broad
wing of amplitude $A/\gamma$ redward of $x_{\rm inj}$ will result as
photons are redshifted across the line centre. The distribution will
cut off sharply at $x>x_{\rm inj}$. For an absorber with $T_a=100$~K
($a\simeq0.00472$), and $x_{\rm esc}=200$, $\gamma_{\rm
crit}\simeq2.8\times10^{-10}$. The solutions for
$\gamma=1.7\times10^{-7}$ with $(x_{\rm inj}, x_{\rm esc})=(100,200)$
and $(25, 50)$ are shown in Fig.~\ref{fig:nxFPDAg}. A substantial
decrease in $n(x)$ is produced compared with the $\gamma=0$ case. A
significant enhancement of the \Lya\ collision rate over the
free-streaming limit, however, may still result, with $P_{l, {\rm
ss}}/P_l(0)\simeq a/(\pi\gamma x_1)$. For $\gamma<0$, a broad wing
blueward of $x_{\rm inj}$ will form as photons are blueshifted in the
contracting gas. A complex profile, however, may form redward of
$x_{\rm inj}$ as redward diffusion is resisted by blueshifting. Such a
case is illustrated in Fig.~\ref{fig:nxFPDAg} for
$\gamma=-1.7\times10^{-7}$ with $x_{\rm inj}=25$ and $x_{\rm esc}=50$.

\begin{figure}
\includegraphics[width=3.3in]{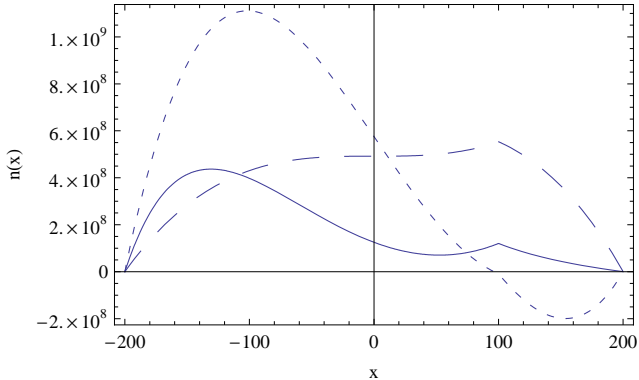}
\caption{The steady-state distribution of \Lya\ resonance line photons
in the diffusion approximation for a source function $\tilde
S(x)=\delta_D(x-x_{\rm inj})$ with $x_{\rm inj}=100$, allowing photons
with $\vert x\vert>x_{\rm esc}=200$ to escape and including the effect
of atomic recoils. Shown are the solution with $\epsilon=0.0079$,
corresponding to $T_a=10$~K (solid line), the solution with
$\epsilon=0$ (long-dashed line), and the first order solution
$n_0(x)+\epsilon n_1(x)$ for $\epsilon=0.0079$ (short-dashed line).
}
\label{fig:nxFPDAe}
\end{figure}

\subsection{Steady-state solution with atomic recoil}
\label{subsec:ssrec}
The above solutions neglect the effect of atomic recoil on the
evolution of the radiation field. Since the photon momentum ($\sim
h_{\rm P}\nu_0/c$) is small compared with the momentum of the hydrogen
atoms ($\sim m_{\rm H}b$), the effect should be small, but may result
in a significant heating rate when the gas is still cold \citep{MMR97,
2004ApJ...602....1C, Meiksin06}. Allowing for atomic recoils modifies
Eq.~(\ref{eq:FPDA}) to
\begin{equation}
\frac{1}{2}\frac{d}{dx}\left[2\epsilon\phi_V(a,x)n(x) +
D(a,x)\frac{dn(x)}{dx}\right] = -\tilde S(x),
\label{eq:FPDAe}
\end{equation}
where the recoil parameter $\epsilon=h_{\rm P}\nu_0/(2k_{\rm B}Tm_{\rm
H}c^2)^{1/2}\simeq0.025T^{-1/2}$ is the ratio of the momentum of a
resonance line photon to the momentum of an atom moving at the thermal
velocity \citep{1994ApJ...427..603R, Meiksin06}. Eq.~(\ref{eq:FPDAe})
expresses only the first order correction in $\epsilon$. The solution
for a source $\tilde S(x)=A\delta_D(x-x_{\rm inj})$ is
\begin{equation}
n(x) = \cases{g_\epsilon(x,-x_{\rm esc}) & ; $-x_{\rm esc}<x<x_{\rm inj}$\cr
g_\epsilon(x,-x_{\rm esc})-g_\epsilon(x,x_{\rm inj})
&; $x_{\rm inj}<x<x_{\rm esc},$\cr}
\label{eq:nxFPDAsse}
\end{equation}
where $g_\epsilon(x,y)=A[\pi/(2\epsilon^3 a)]\exp(-2\epsilon x)
[f_\epsilon(x)-f_\epsilon(y)] [f_\epsilon(x_{\rm
esc})-f_\epsilon(x_{\rm inj})]/ [f_\epsilon(x_{\rm
esc})-f_\epsilon(y)]$, and $f_\epsilon(x)=(1-2\epsilon x + 2\epsilon^2
x^2)\exp(2\epsilon x).$ Here, the wing approximation for $\phi_V(a,x)$
is adopted, but we find a nearly identical result allowing for the
Doppler core as well.

The solution for $x_{\rm inj}=100$, $x_{\rm esc}=200$ and
$\epsilon=0.0079$, corresponding to $T\simeq10$~K, is shown in
Fig.~\ref{fig:nxFPDAe}. The result differs substantially from the
solution for $\epsilon=0$. In fact the solution cannot be trusted, as
it is not consistent with the original order of Eq.~(\ref{eq:FPDAe})
in $\epsilon$. This may be demonstrated by adding to
$2\epsilon\phi_V(a,x)n(x)$ the quantity $\epsilon^2 y(x)$, where
$y(x)$ is an unspecified function representing the higher order
corrections to the recoil, and which in general may also depend on
$n(x)$. Inserting the series solution
$n(x)=\Sigma_{i=0}^\infty\,\epsilon^i n_i(x)$ into
Eq.~(\ref{eq:FPDAe}) and equating equal orders in $\epsilon$ recovers
$n_0(x)$ as given by Eq.~(\ref{eq:nxFPDAss}), and produces the first
order equation $(d/dx)[2\phi_V(a,x)n_0(x)+D(a,x)dn_1(x)/dx]=0$ and the
second order equation
$(d/dx)[2\phi_V(a,x)n_1(x)+y(x)+D(a,x)dn_2(x)/dx]=0$. The first order
equation gives $n_1(x)$ in terms of $n_0(x)$, subject to the boundary
conditions $n_1(-x_{\rm esc})=n_1(x_{\rm esc})=0$.  The second order
equation shows that $n_2(x)$ depends on the unspecified (neglected)
function $y(x)$. Thus the solution is valid only up to $n(x) = n_0(x)
+ \epsilon n_1(x)$. If higher order terms are substantial, then the
solution for $n(x)$ cannot be trusted. The solution $n_0(x)+ \epsilon
n_1(x)$ for the example above differs substantially from the solution
$n(x)$, and indeed is not even positive-definite, as shown in
Fig.~\ref{fig:nxFPDAe}, demonstrating the solution is not valid.

Including higher order terms involves a non-trivial expression for the
redistribution function including non-linear terms in the recoil
parameter \citep{1981Ap.....17...69B, Meiksin06}. (Additional higher
order terms in the recoil parameter arise from relativistic
corrections, but these are reduced by factors of $b/c$.) For
temperatures $T>100$~K, the discrepancy between the full solution and
the first order solution is smaller than about 20\% for $x_{\rm
inj}=100$ and $x_{\rm esc}=200$. In general, the correction will be
small for $\epsilon x_{\rm esc}\ll1$, and we caution that if this is
violated the steady-state solutions may not be valid. We do not pursue
further consequences of the atomic recoil solution here.

\subsection{The role of bulk peculiar velocities}
\label{subsec:vpec}

Absorption systems with a bulk peculiar velocity component away from
the source will see the photons redshifted into line centre. In this
case, the scattering rate of resonance line photons may be appreciable
even in an optically thin absorber ($N_{19}<N_{19, {\rm crit}}$), so
that the build-up of a strong radiation field at line centre through
photon capture and diffusion is not necessary for significant WFE
decoupling of the spin state from the CMB. The peculiar velocity
required to bring the absorber into the peak of the source function
$S(\nu)$ in Eq.~(\ref{eq:SnuLnu}) is given by $v_{\rm pec}\simeq
\frac{1}{2}x_1 b_a$, where $x_1$ is given by Eq.~(\ref{eq:x1}). For
pure thermal broadening (assuming negligible internal velocity
structure within the absorber), this corresponds to
\begin{equation}
v_{\rm pec}\simeq14h^{-1}(1+z)^{3/2}
\left(\frac{T_a}{T_{\rm IGM}}\right)^{1/2}\kms.
\label{eq:vpec}
\end{equation}
This is comparable to the typical peculiar velocities of non-linear
cosmological structures at the epochs of interest. As a result, {\it
the WFE will produce a 21cm signature that maps the peculiar velocity
field of the structures within the IGM}.

\begin{figure}
\includegraphics[width=3.3in]{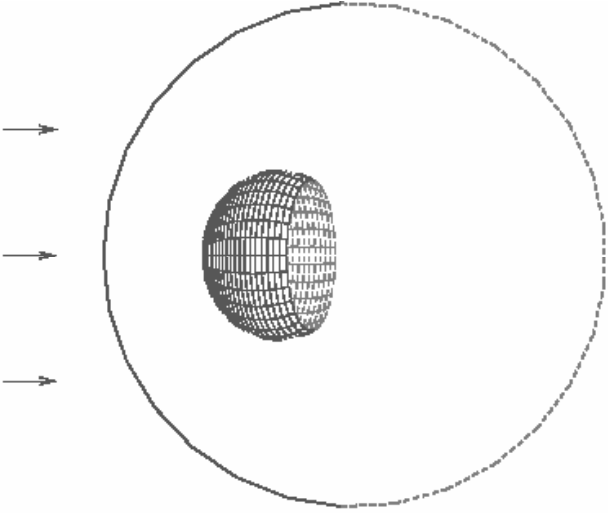}
\caption{Surface of constant \Lya scattering rate $P_l$ produced by a
continuum source to the left within a collapsing halo of neutral
hydrogen. The inner surface terminates at the accretion shock. The
circle represents the turnaround radius $r_{\rm ta}$ of the halo, with
the solid line representing the hemisphere of gas with peculiar
velocity redshifted relative to the source and the dashed line the
hemisphere with blueshifted peculiar velocity. The surface shown
corresponds to a maximum radius of 0.59$r_{\rm ta}$. The figure is
tilted by $5^\circ$ to reveal the 3D structure of the surface.
}
\label{fig:accretion}
\end{figure}

As an illustration, we consider the signal produced by a collapsing
halo, such as around a forming galaxy or galaxy cluster, illuminated
by an external radiation source, as depicted in
Fig.~\ref{fig:accretion}.

The collision rate from a source of specific luminosity $L_\nu$ on an
absorber with Doppler parameter $b_a$ at luminosity distance $r_L$
moving away from the source with peculiar velocity $v$ is
\begin{eqnarray}
P_l&=&\sigma\frac{L_\nu}{h_{\rm P}\nu}\frac{1}{4\pi r_L^2}
\int_0^\infty d\nu\,
\varphi_V\left[a,\nu(1-\frac{v}{c})\right]\exp(-\tau_\nu)\nonumber\\
&\simeq&\sigma\frac{L_\nu}{h_{\rm P}\nu}\frac{1}{4\pi r_L^2}
\exp\left[-\frac{b_a x_1}{v_0\cos(\theta)}\right],
\label{eq:Placc}
\end{eqnarray}
using Eq.~({\ref{eq:taunu_ap}}), approximating
$\phi_V(a,x)\simeq\pi^{-1/2}\exp(-x^2)$, noting that the resulting
exponential in the integrand peaks sharply near $x=v/b_a$ assuming
$v_0/b_a\simeq x_1$ or larger, Taylor expanding the argument of the
exponential to second order in $x-v/b_a$, and expressing the peculiar
velocity as $v=v_0\cos(\theta)$, the projected accretion velocity
along the line of sight to the source.

In general, $v_0$ will vary with distance from the centre of the
collapsing halo. As a definite case, we approximate the peculiar
velocity of the accreting gas using the self-similar solution of
\citet{1985ApJS...58...39B} for adiabatic accretion of a $\gamma=5/3$
gas of negligible density onto a dark matter halo with a comoving
centre in an Einstein-de Sitter cosmology. The velocity field in the
region between the accretion shock and the turn-around radius at
cosmological time $t$ is well-approximated by
\begin{equation}
v_0(r)\simeq\frac{r_{\rm ta}}{t}\frac{V_{\rm min}}{\log_{10}\lambda_s}
\log_{10}\lambda,
\label{eq:vacc}
\end{equation}
where $r_{\rm ta}$ is the turn-around radius, $\lambda=r/r_{\rm ta}$,
$\lambda_s=r_s/r_{\rm ta}\simeq0.3472$, where $r_s$ is the radius of the
accretion shock, and $V_{\rm min}\simeq-1.433$ characterises the inflow
velocity of the gas just before passing through the accretion shock.

Contours of constant values of $P_l$ correspond to values of constant
projected accretion velocity. If at a radius $r$ towards the source
($\theta=0$), the accretion velocity is a fraction $f$ of its maximum
infall velocity, then contours of fixed $P_l$ will correspond to the
surfaces
\begin{equation}
\lambda=\lambda_s^{f\sec(\theta)},
\label{eq:Plcont}
\end{equation}
extending over the angular range $0\leq\theta\leq \arccos(f)$. The
surface corresponding to $f=0.5$ is shown in Fig.~\ref{fig:accretion},
with a maximum angle $\theta_{\rm max}=60^\circ$ and maximum radius
$\lambda_0=\lambda_s^f=0.59$. The loci of constant $P_l$ correspond to
arcing sheets. The exponential sensitivity of $P_l$ to $v$ for
$v_0/b_a \simeq x_1$ (see Eq.~[\ref{eq:Placc}]) ensures a surface of
narrow width will dominate the 21cm signal produced. Such a surface
could be mistaken for the region of neutral hydrogen just outside an
\HII region surrouding a source of ionizing radiation. A varying gas
temperature, and therefore varying $b_a$, in the accreting region will
further complicate the signal. If $v_0/b_a>>x_1$, then the entire
solid hemisphere will be illuminated by the source (except for a thin
circular wafer at $\theta\la90^\circ$).

A similar effect may apply to an expanding void. Since voids expand
faster than the Hubble expansion, the peculiar velocity on the far
side of a void from a distant source could redshift the received
radiation into line centre. A shell of direct scattering may then
result even without including the reddening effects of radiative
transfer within the void.

\section{The evolution of the WFE in discrete objects}
\label{sec:evol}

It was assumed for the solutions in \S~\ref{sec:discrete} that the
photons are able to achieve a steady state as they scatter within the
absorbing system. In this section, we compute the timescales required
to establish a steady state by solving the time-dependent diffusion
equation
\begin{equation}
\frac{\partial n(x,\tau)}{\partial\tau}=\frac{1}{2}\frac{\partial}{\partial x}
\left[D(a,x)\frac{\partial n(x,\tau)}{\partial x}\right]+\tilde S(x),
\label{eq:FPDAt}
\end{equation}
where $\tau=t/t_s$, and $t_s=\Delta\nu_D/(n_l\sigma
c)\simeq3.2n_l^{-1}T_a^{1/2}$~s is the scattering time of the
resonance line photons at line centre.

\begin{figure}
\includegraphics[width=3.3in]{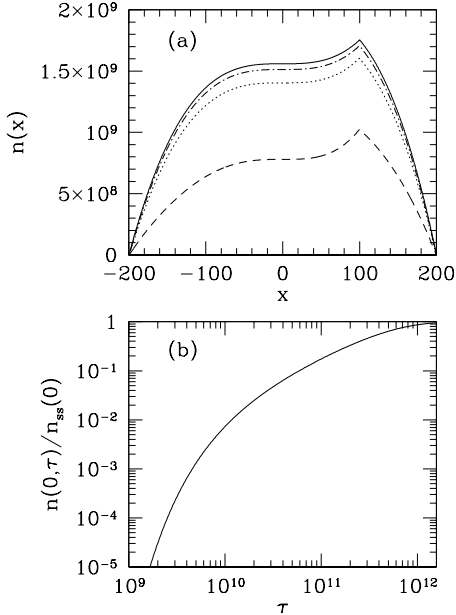}
\caption{(a)\ Convergence of the time-dependent solution to the
diffusion equation for a source $\tilde S(x)=\delta_D(x-x_{\rm inj})$
with $x_{\rm inj}=100$, $x_{\rm esc}=200$ and $T_a=100$~K. Shown are
the steady-state photon frequency distribution,
Eq.~(\ref{eq:nxFPDAss}) (solid line), and the photon frequency
distributions at the times their values at $x=0$ take on 50\% (dashed
line), 90\% (dotted line) and 97\% (dot-dashed line) of the
steady-state value. (b)\ The evolution of the ratio of the photon
frequency distribution at line centre ($x=0$) to the steady-state
value, as a function of time in units of the scattering time at line
centre.
}
\label{fig:nxFPDAt}
\end{figure}

\begin{figure}
\includegraphics[width=3.3in]{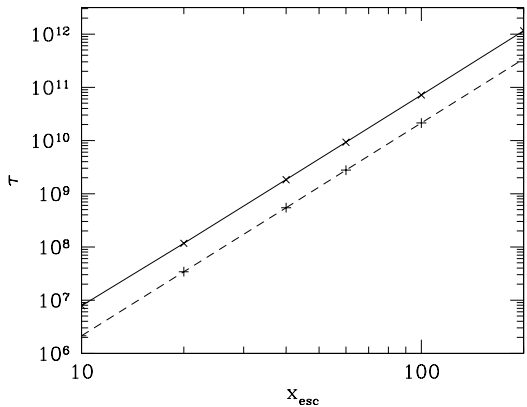}
\caption{The convergence time in units of the scattering time at line
centre for the photon frequency distribution with source $\tilde
S(x)=\delta_D(x-x_{\rm inj})$ having $x_{\rm inj}=x_{\rm esc}/2$ to
reach 50\% (dashed line) and 90\% (solid line) of the steady-state
value at $x=0$. A temperature of $T_a=100$~K is assumed.
}
\label{fig:convergencetime}
\end{figure}

We solve Eq.~(\ref{eq:FPDAt}) using a Crank-Nicholson scheme. We
tested the scheme against the time-dependent solutions for a constant
$\delta$-function source at line centre in an infinite homogeneous
medium and a flat continuum source in the diffusion approximation
\citep{1994ApJ...427..603R}. The results applied to the source $\tilde
S(x)=\delta_D(x-x_{\rm inj})$ with $x_{\rm inj}=100$ and boundary
condition $n(x)=0$ for $\vert x\vert>x_{\rm esc}=200$ are shown in the
upper panel of Fig.~\ref{fig:nxFPDAt}. The photon frequency
distribution takes on the approximate shape of the final steady-state
solution Eq.~(\ref{eq:nxFPDAss}) as it converges towards it. The
evolution of the photon frequency distribution at line centre ($x=0$)
is shown in the lower panel of Fig.~\ref{fig:nxFPDAt}. For sources
that have been active only a small fraction of the time required to
achieve full convergence, the scattering rate may be negligibly small.

Dimensional analysis of Eq.~(\ref{eq:FPDAt}) suggests the time to
establish a steady state for photons injected at $x=x_{\rm inj}$ is
$\tau_{ss}\simeq2x_{\rm inj}^2/D(a,x_{\rm inj})\simeq(2\pi/a)x_{\rm
inj}^4 =(\pi/8a)x_{\rm esc}^4\simeq8.3T_a^{1/2}x_{\rm esc}^4$ for
$x_{\rm inj}=x_{\rm esc}/2$. The time for the photon distribution to
converge to the steady-state value at $x=0$ found from the numerical
integration of Eq.~(\ref{eq:FPDAt}) is shown in
Fig.~\ref{fig:convergencetime} for convergence to 50\% and 90\% of the
steady-state value, for a source $\tilde S(x)=\delta_D(x-x_{\rm inj})$
with $x_{\rm inj}=x_{\rm esc}/2$ and $T_a=100$~K. The convergence
times scale like $\tau_{ss, 50}\simeq210 x_{\rm esc}^4$ (50\%) and
$\tau_{ss, 90}\simeq750 x_{\rm esc}^4$ (90\%), in good agreement with
the expected behaviour.

\section{Discussion and conclusions}
\label{sec:disc}

The role played by the Wouthuysen-Field effect in decoupling the spin
temperature of the neutral hydrogen from the CMB temperature depends
most crucially on the scattering rate of \Lya\ photons. The spin
temperature $T_S$ is generally a weighted mean of the colour
temperature $T_\alpha$, kinetic temperature $T_K$ of the gas, and
brightness temperature $T_R$ of any incident radiation (such as the
CMB) at the 21cm frequency,
\begin{equation}
T_S = \frac{T_R + y_\alpha T_\alpha + y_c T_K}{1+y_\alpha+y_c},
\label{eq:TS}
\end{equation}
where
\begin{equation}
y_\alpha\equiv\frac{P_{10}}{A_{10}}\frac{T_*}{T_\alpha}\qquad{\rm and}\qquad
y_c\equiv\frac{C_{10}}{A_{10}}\frac{T_*}{T_K}
\label{eq:weights}
\end{equation}
are the \Lya\ and collisional pumping efficiencies, respectively
\citep{1958PROCIRE.46..240F, MMR97}, $A_{10}=2.85\times10^{-15}\,{\rm
s}^{-1}$ is the spontaneous decay rate of the 21cm transition, and
$T_*\equiv h_{\rm P}\nu_{10}/k_{\rm B}$, where $\nu_{10}$ is the
frequency of the 21cm transition. Here $P_{10}= 4P_l/27$ is the
indirect de-excitation rate of the triplet hyperfine state induced by
\Lya\ photon scattering. The colour temperature of the radiation field
is the harmonic mean temperature $T_\alpha=1/\langle
T_u^{-1}(\nu)\rangle$, where $T_u(\nu)=-(h_{\rm P}/k_{\rm B})(d\ln
u_\nu/d\nu)^{-1}$, weighted by $u_\nu \varphi_V(a,\nu)$ for an energy
density $u_\nu$ of resonance line photons \citep{Meiksin06}. The
coefficient $C_{10}$ is the de-excitation rate of the triplet
hyperfine state induced by atomic collisions.

For a \Lya\ scattering rate $P_l$ exceeding the critical
thermalization rate
\begin{equation}
P_{\rm th}=\frac{27 A_{10}T_{\rm CMB}}{4T_*}
\simeq6.8\times10^{-12}\,{\rm s}^{-1}\left(\frac{1+z}{9}\right),
\label{eq:Pth}
\end{equation}
where $T_{\rm CMB}\simeq2.73(1+z)$
\citep{Mather94} is the CMB temperature, the spin temperature will be
driven to the colour temperature in the absence of strong collisional
de-excitation \citep{MMR97}. The radiation field will typically
rapidly thermalize with the neutral hydrogen, resulting in
$T_\alpha\rightarrow T_K$ \citep{1959ApJ...129..551F, Meiksin06}. The
neutral hydrogen will then appear in either absorption or emission
against the CMB, depending on whether $T_K$ is below or above $T_{\rm
CMB}$, respectively.

Before the Epoch of Reionization, when the first radiation sources
turn on at the End of the Dark Ages, regions around the sources may
become visible through their 21cm signature. Because the photons
produced by a source, such as a QSO or star-forming galaxy, near the
\Lya\ resonance will be scattered out of the line of sight over short
distances, the scattering rate of \Lya\ photons will be much smaller
than $P_{\rm th}$ except extremely near the source. Only photons
emitted sufficiently blueward of the \Lya\ resonance frequency will
survive to any distant structures, where they will be received well in
the blue Lorentz wing. If a structure has a sufficient optical depth
in the wing to the photons received, it will trap them and
rescatterings within the structure will allow the photons to diffuse
in frequency across the resonance line. The radiation field may be
built up in this way until it reaches a steady state in which the rate
of incoming photons is balanced by the rate at which photons diffuse
sufficiently far from the resonance line frequency to escape in the
wings where the structure is optically thin. The resulting photon
scattering rate in the diffusion approximation may greatly exceed the
estimate $P_l(0)$ in the free-streaming limit, for which it is assumed
no resonance line photons emitted by the source have been scattered
out of the line of sight before reaching the absorber.

We identify several factors, however, that may reduce the energy
density of the resonance line photons from the steady-state diffusion
approximation estimate for static structures:

(1) Internal expansion within the structure, characterised by the
dimensionless expansion parameter $\gamma$, will redshift the photons
into a red wing, where they will escape, on a timescale much faster
than the diffusion time, resulting in a much reduced energy
density. In the diffusion approximation, the scattering rate will be
reduced compared with the $\gamma=0$ case by the factor
$(3a/\pi\gamma)/(x_{\rm esc}^3-x_{\rm inj}^3)$ for
$\gamma\gg\gamma_{\rm crit}=3a/(2\pi x_{\rm esc}^3)$, where $x_{\rm
  esc}$ is the dimensionless escape frequency at which photons escape
the structure, and $x_{\rm inj}$ is the frequency at which photons are
predominantly received from the source. Contraction within the
structure may altogether prevent photons from diffusing across the
resonance line frequency.

(2) The scattering rate may be reduced by atomic recoil. Estimating
the magnitude of this effect, however, generally requires including
non-trivial higher order terms in the recoil parameter $\epsilon$ than
first order in the steady-state equation. We caution against
introducing atomic recoils into Monte Carlo calculations and other
steady-state solutions, as the results obtained may not be physically
self-consistent if the solutions differ substantially from those
obtained for $\epsilon=0$.

(3) The time scale to establish a steady state will typically exceed
the duration of high star formation rates in galaxies or the lifetime
of QSOs. As a consequence, the collision rate will be reduced by
factors of tens to thousands compared with the steady-state value.

As an illustration, consider the region around a bright QSO with
specific luminosity at the Lyman edge of $10^{31}\,{\rm
ergs\,s^{-1}\,Hz^{-1}}$ pre-heated by soft x-rays ahead of the
ionization front to a mean temperature of $T_{\rm IGM}\simeq1800$~K at
a distance of 10~Mpc \citep{MMR97}. According to Eq.~(\ref{eq:x1}),
$x_1\simeq200$ at $z=8$ (assuming $h=0.7$). A structure 250~kpc across
with a moderate overdensity of 3 and internal temperature of
$T_a\simeq120$~K will have a neutral hydrogen column density of
$N_{\rm HI}\simeq3\times10^{20}\,{\rm cm^{-2}}$ and escape frequency
$x_{\rm esc}\simeq200$, from Eq.~(\ref{eq:xesc}). For a power law
spectrum $\nu^{-1.5}$, Eq.~(\ref{eq:Pl0}) gives for the \Lya\
collision rate in the free-streaming limit
$P_l(0)\simeq8.7\times10^{-13}\,{\rm s^{-1}}$, about an order of
magnitude smaller than the thermalization rate $P_{\rm th}$. The
steady-state value from Eq.~(\ref{eq:Plratioss}) then gives $P_{l,
{\rm ss}}\simeq10^{-8}\,{\rm s^{-1}}$, well in excess of $P_{\rm
th}$. The time to establish a steady state, however, is long. The
scattering time of \Lya\ photons in the structure is about
$t_s\simeq8.5\times10^4$~s. According to
Fig.~\ref{fig:convergencetime}, the time for the radiation field to
achieve a steady state is then about $10^{17}$~s, or 3~Gyr, greatly
exceeding the expected lifetime of the QSO. If the QSO had illuminated
the structure for as long as 0.1~Gyr, according to
Fig.~\ref{fig:nxFPDAt} the steady-state value must be reduced by a
factor of $0.05$. The resulting scattering rate will still exceed
$P_{\rm th}$. If the system has a peculiar velocity away from the QSO,
the rate could be quite different. Eq.~(\ref{eq:vpec}) shows that a
peculiar velocity of $140\kms$, a plausible value, would shift the
peak of the radiation from the source received by the structure to the
resonance line frequency. The radiation field would then rapidly reach
its steady-state value, and $P_{l, {\rm ss}}$ would greatly exceed
$P_{\rm th}$.

A much more overdense structure at the same distance from the QSO
would be able to achieve the steady-state scattering rate. For
example, a virialized structure 35~kpc across with an overdensity of
200 and temperature $T_a\simeq1000$~K would have a neutral hydrogen
column density of $3\times10^{21}\,{\rm cm^{-2}}$ and $x_{\rm
esc}\simeq220$. The time for the radiation field to reach a steady
state would then be about 0.2~Gyr, a plausible lifetime for the QSO.

By contrast, it may be difficult to achieve a steady state in a void.
For an IGM temperature $T_{\rm IGM}\simeq100$~K, $x_1=825$ at $z=8$.
A region 3~Mpc across with a temperature of $T_a=10$~K, and underdense
by a factor of 3 will have a neutral hydrogen column density of
$4\times10^{20}\,{\rm cm^{-2}}$, $x_{\rm esc}\simeq800$ and a
scattering time at line centre of $t_s\simeq2.2\times10^5$~s. The time
to reach a steady state is then 2100~Gyr. Even if the void is
illuminated for as long as the age of the Universe, the radiation
field will be negligibly small compared with its steady-state
value. Since the void will expand, however, the incident radiation
will be redshifted across the resonance
line. \citet{1994ApJ...427..603R} estimate a characteristic time for
redshifting to dominate diffusion in an infinite homogeneous medium to
be $\tau_{\gamma}\simeq(a/\gamma^4)^{1/3}$. Voids expand somewhat
faster than the mean Hubble expansion. Adopting the Hubble constant at
$z=8$ for $h=0.7$, $\gamma\simeq5\times10^{-6}$. A steady state should
be established for $t\gg\tau_{\gamma}t_s\simeq2\times10^4$~yr, and the
steady-state scattering rate will then match the free-streaming value,
with $P_{l, {\rm ss}}/P_l(0)\simeq a/(\pi\gamma x_1) \simeq1$, which
may be adequate for producing a 21cm signature in the presence of
sufficient sources.

The examples above show that the effect of peculiar motions may lead
to an enhancement in the \Lya\ scattering rate over the free-streaming
limit and produce the required decoupling of $T_S$ from $T_{\rm
CMB}$. Even if the scattering rate is inadequate for inducing
decoupling throughout most of the structure, it may be adequate along
narrow loci within which the velocity field either permits the
resonance line photons to accumulate near the resonance line frequency
or shifts the resonance line frequency of the absorber to matching the
incident radiation field from the source as filtered through the
IGM. We presented an example of an accreting halo which produces a
21cm signature along a curved surface for which the impinging photons
are received at the resonance line frequency due to the infall
velocity of the accreting gas. Such a curved structure could be
mistaken for a rim of neutral gas outside an \HII\ region in radio
maps of the IGM. Similar curved surfaces may form on the far side of
voids from a distant source.

As a consequence of the above effects, the 21cm signature of the IGM
will not simply trace the density structure of the IGM, but will trace
the peculiar velocity structure as well. We have not considered
multiple sources. In the presence of a large number of nearby sources,
a greater fraction of the gas will be able to match the incident
radiation field through their peculiar motions. In addition, the
escaping radiation from the absorbers will contribute to the ambient
radiation field. In due course, a metagalactic diffuse radiation field
may grow sufficiently strong to ensure the WFE is capable of
decoupling the spin temperature from the CMB temperature throughout
the IGM, and so ensure the production of a 21cm signature
everywhere. The effects described here suggest that in the early
stages, before any such metagalactic field has been established (and
it has yet to be demonstrated that such a field will be established),
the 21cm signatures produced at the End of the Dark Ages, and possibly
extending into the Epoch of Reionization, may only be interpreted with
a thorough understanding of the evolution of the energy density of
resonance line photons within a clumpy medium, including the effects
of peculiar motions.

\begin{appendix}
\section{Optical depth through Doppler core}
\label{ap:taucore}

Radiation from a source at $z_S$ received at $z$ with frequency $\nu$
in the range $\nu_0[(1+z)/(1+z_S)]<\nu<\nu_0$, where $\nu_0$ is the
resonance line frequency, will have passed through the resonance line
frequency en route to the absorber. To estimate the optical depth of
the IGM at these frequencies, it is convenient to approximate the
Voigt line profile as
\begin{eqnarray}
\phi_V(a,x)&=&\pi^{-1/2}\exp(-x^2);\qquad -x_m < x < x_m\nonumber\\
&=&\frac{a}{\pi x^2};\qquad \vert x\vert > x_m
\label{eq:phiVap}
\end{eqnarray}
where $x_m$ is the value of $x=(\nu-\nu_0)/\Delta\nu_D$ at which the
two approximations match. For $T=10$~K, 100~K and 1000~K, the values of
$x_m$ are 2.58, 2.83 and 3.05, respectively.

A photon of frequency $\nu$ at $z$ will have $x=\pm x_m$ at the
redshift $z_m^{(\pm)}$, given by
\begin{equation}
\frac{\nu}{\nu_0}\frac{1+z_m^{(\pm)}}{1+z}=1\pm x_m\frac{b}{c}
\label{eq:zm}
\end{equation}
where $b=(2k_{\rm B}T_{\rm IGM}/m_{\rm H})^{1/2}$ is the Doppler
parameter of the IGM at temperature $T_{\rm IGM}$, assumed
constant. (In principle, a turbulent velocity component may be added
in quadrature.) The optical depth in Eq.~(\ref{eq:taunu}) may then be
expressed as a sum of three contributions, corresponding to the ranges
$x<-x_m$, $-x_m<x<x_m$ and $x>x_m$,
\begin{eqnarray}
\tau_\nu&=&\frac{\sigma a}{\pi\Delta\nu_D}\int_z^{z_m^{(-)}} dz^\prime\,
\frac{dl_p}{dz^\prime}n_l(z^\prime)\frac{(\Delta\nu_D)^2}
{\left(\nu\frac{1+z^\prime}{1+z}-\nu_0\right)^2}\nonumber\\
&+&\frac{\sigma}{\pi^{1/2}\Delta\nu_D}\int_{z_m^{(-)}}^{z_m^{(+)}}
dz^\prime\, \frac{dl_p}{dz^\prime}n_l(z^\prime)\exp\left[-\left(\frac
{\nu\frac{1+z^\prime}{1+z}-\nu_0}
{\Delta\nu_D}\right)^2\right]\nonumber\\
&+&\frac{\sigma a}{\pi\Delta\nu_D}\int_{z_m^{(+)}}^{z_S} dz^\prime\,
\frac{dl_p}{dz^\prime}n_l(z^\prime)\frac{(\Delta\nu_D)^2}
{\left(\nu\frac{1+z^\prime}{1+z}-\nu_0\right)^2}.
\label{eq:taunufull}
\end{eqnarray}
The first and third terms are similar to Eq.~(\ref{eq:taununo}).
The dominant term is the second, representing scattering through the line core.
Setting $u=[(\nu(1+z^\prime)/(1+z)-\nu_0)/\Delta\nu_D]^2$, and
assuming $b/c\ll1$ within the integrand, gives for this contribution
\begin{eqnarray}
\tau_\nu^{\rm core}&\simeq&\frac{1}{\pi^{1/2}}\frac{\sigma cn(0)}
{H_0\Omega_m^{1/2}}(1+z)^{3/2}\nu_0^{1/2}\nu^{-3/2}\gamma(\frac{1}{2},x_m^2)
\nonumber\\
&\simeq&8000h^{-1}(1+z)^{3/2}y^{-3/2}\gamma(\frac{1}{2},x_m^2),
\label{eq:taunucore}
\end{eqnarray}
where $y=\nu/\nu_0$, $\gamma(t,x)=\int_0^x du u^{t-1}\exp(-u)$ is an
incomplete gamma function, and $\Omega_m=0.3$ and $\Omega_bh^2=0.022$
are assumed. This will in general ensure that essentially no flux from
the source is received in the frequency range
$\nu_0(1+z)/(1+z_S)<\nu<\nu_0$. This is the Gunn-Peterson effect.

\section{Slab solution}
\label{ap:slab}

The diffusion equation for the transfer of radiation through a static
slab in the Eddington approximation is given by
\cite{1973MNRAS.162...43H} as
\begin{eqnarray}
\frac{\partial^2n_x(\tau,\sigma)}{\partial\tau^2}
+\frac{\partial^2n_x(\tau,\sigma)}{\partial\sigma^2}&=&
-3\phi_V^2(a,x)\frac{G(\tau)}{4\pi},\nonumber\\
&\simeq&-6^{1/2}\delta_D(\sigma)\frac{G(\tau)}{4\pi},
\label{eq:HDAslab}
\end{eqnarray}
for a spectrally flat source $G(\tau)$, in the limit
$\phi_V(a,x)\simeq a/(\pi x^2)$. Here, $\sigma$ is related to $x$
through $dx/d\sigma=(3/2)^{1/2}\phi_V(a,x)$, and
$\tau=\tau_\nu/\phi_V(a,x)$ is the mean optical depth vertically
through the slab. The photon number density $n(x)$ is expressed as
$n_x(\sigma)$. The last line in Eq.~(\ref{eq:HDAslab}) follows from
approximating $3\phi_V^2(a,x)$ as a Dirac $\delta$-function,
$6^{1/2}\delta_D(\sigma)$ where the coefficient preserves the
normalization. The boundary conditions assumed are
\begin{equation}
\frac{\partial n_x(\tau,\sigma)}{\partial\tau}{\Bigg\vert}_{\pm B}=
\mp\frac{3}{2}\phi_V(a,x)n_x(B,\sigma)
\label{eq:HDAslab-bct}
\end{equation}
and
\begin{equation}
n_x(\tau,\sigma)\rightarrow0\quad {\rm for} \quad\sigma\rightarrow\pm\infty.
\label{eq:HDAslab-bcs}
\end{equation}

For a uniformly distributed source of unit strength,
\cite{1973MNRAS.162...43H} takes $G=1/(2B)$, where $-B<\tau<B$
describes the vertical extent of the slab. In terms of the total
optical depth at line centre, $\tau_0=2B\phi_V(a,0)=2B/\pi^{1/2}$, the
source becomes $G=\pi^{-1/2}/\tau_0$. The solution at the centre of
the slab is
\begin{equation}
n_x(0,x)=\frac{6^{1/2}}{2\pi^3}S\left[\exp\left(-2^{1/2}\left(\frac{\pi}{3}
\right)^{3/2}\frac{\vert x\vert^3}{a\tau_0}\right)\right],
\label{eq:HDAslab-sol}
\end{equation}
where the function $S(z)=z-z^3/3^2+z^5/5^2-+\dots\simeq z$ for $0<z<1$.

The diffusion approximation Eq.~(\ref{eq:FPDA}) transforms to
\begin{equation}
\frac{d^2n_x(\sigma)}{d\sigma^2}=-6^{1/2}A\delta_D(\sigma),
\label{eq:FPDAslab}
\end{equation}
for a source $\tilde S(x)=A\delta_D(x)$ and adopting
$D(a,x)=\phi_V(a,x)$. Identifying $A=G/(4\pi)=1/(4\pi^{3/2}\tau_0)$
reproduces Eq.~(\ref{eq:HDAslab}) with the $\tau$-dependence
suppressed. Expressing $x_{\rm esc}=x_*(a\tau_0)^{1/3}$, the solution
Eq.~(\ref{eq:nxFPDAss}) becomes
\begin{equation}
n_x(x) =\frac{x_*^3}{12\pi^{1/2}}\left(1-\frac{\vert
x\vert^3}{x_*^3a\tau_0}\right) \qquad(\vert x\vert\leq x_{\rm esc}),
\label{eq:FPDAslab-sol}
\end{equation}
for $x_{\rm inj}=0$. Noting $S(1)\simeq0.92$, equating the values of
Eqs.~(\ref{eq:HDAslab-sol}) and (\ref{eq:FPDAslab-sol}) at $x=0$ gives
$x_*\simeq0.92$. The two profiles are very similar, with
Eq.~(\ref{eq:HDAslab-sol}) forming an exponential tail with a cutoff
at $x\gta0.87(a\tau_0)^{1/3}\simeq0.95x_{\rm esc}$.

\end{appendix}



\bibliographystyle{mn2e-eprint}
\bibliography{apj-jour,wfr}

\label{lastpage}

\end{document}